\documentclass[10pt, a4paper]{article}

\usepackage{lrec-coling2024} 
\usepackage{algpseudocode}
\usepackage{algorithm}
\usepackage{makecell}

\title{ByteComposer: a Human-like Melody Composition Method based on Language Model Agent}
\name{Xia Liang, Xingjian Du, Jiaju Lin, Pei Zou, Yuan Wan, Bilei Zhu}

\address{}

\address{ByteDance AI Lab \\
         Shanghai, China \\
         \{liangxia.21, duxingjian.real, zoupei, wanyuan.0626, zhubilei\}@bytedance.com\\
         jjl7137@psu.edu}

\abstract{
Large Language Models (LLM) have shown encouraging progress in multimodal understanding and generation tasks. However, how to design a human-aligned and interpretable melody composition system is still under-explored. To solve this problem, we propose ByteComposer, an agent framework emulating a human's creative pipeline in four separate steps : 
"Conception Analysis - Draft Composition - Self-Evaluation and Modification - Aesthetic Selection". This framework seamlessly blends the interactive and knowledge-understanding features of LLMs with existing symbolic music generation models, thereby achieving a melody composition agent comparable to human creators. We conduct extensive experiments on GPT4 and several open-source large language models, which substantiate our framework's effectiveness. Furthermore, professional music composers were engaged in multi-dimensional evaluations, the final results demonstrated that across various facets of music composition, ByteComposer agent attains the level of a novice melody composer.
 \\ \newline \Keywords{Symbolic Melody Generation, Large Language Model, Agent} }

\begin{document}
\maketitleabstract

\section{Introduction}
\label{sec:intro}
With the development of Transformer architecture and language model, text-to-music generation\cite{schneider2023mo, huang2023noise2music, zhu2023ernie, agostinelli2023musiclm, lam2023efficient} attracts more researchers' interests. 
Different from traditional unconditional music generation models, which are characterized by a lack of interactivity and controllability as well as limited utility, text-based music generation methods offer an enticting solution by providing an interface where input text serves as control conditions for generation.

Powered by its strong natural language understanding and sequence modeling ability, transformer-based language models have become an inevitable module in most existing text-to-music methods. Current methods can be broadly divided into two categories: text-to-audio generation and text-to-symbolic generation. 
Text-to-audio models generate music audio end-to-end based on input text, either trained with paired text-audio data \cite{schneider2023mo, huang2023noise2music, zhu2023ernie, JEN-1}, or using a joint text-audio embedding space\cite{huang2023noise2music, agostinelli2023musiclm, lam2023efficient}. However, output in audio form is difficult to modify according to specific user requirements. Text-to-symbolic methods render symbolic output to facilitate the post-process and further modification. Mainstream text-to-symbolic generation models follow a text-attribute-music paradigm, where textual inputs are translated into musically informed attributes for symbolic generation.
Butter \cite{zhang2020butter} defines its text input with three musical attributes: key, meter and style, then applies a variational autoencoder for generation. FiGARO \cite{von2022figaro} utilizes a complex text input design that encompasses three types of musical attributes: instrumentation, harmony, and meta-information. This intricate design has been specifically tailored for music experts, ensuring that they can leverage their knowledge and expertise to create customized music compositions effectively.
MuseCoco \cite{lu2023musecoco}, an advanced text-to-symbolic model, expands the set of musical attributes to a comprehensive extent, covering a total of 12 musical attributes encompassing both subjective and objective aspects. Thus, MuseCoco grants users finer control over the music generated, allowing them to create more diverse and expressive compositions.

Despite the notable achievements in the current state of the art, several challenges persist within the field of text-to-music generation. Firstly, while these methodologies empower users to adjust the generated music by configuring corresponding attributes, they presuppose a certain level of musical proficiency to comprehend these chosen attributes fully. This prerequisite may, therefore, restrict the accessibility of these models for certain users.  Secondly, due to the substantial cost associated with annotating symbolic data, text-to-symbolic methods face issues related to data scarcity, rendering it difficult to generalize to attributes that have not been previously encountered. Thirdly, the generation process remains black-box, lacking both explainability and fine-grained control. 

To address these problems, we propose ByteComposer, an LLM-driven melody composer with human-like composition procedures. Particularly, to enhance interactivity, we employ an LLM as a music expert to bridge the gap between common user queries and attributes for music generation. By seamlessly mapping users' natural language queries to musical attributes, ByteComposr is capable of understanding common users' intentions and further expanding its usage scenarios. 
Concurrently, through sophisticated prompt engineering, we can extract professional music knowledge from the LLM, thereby enabling adaptation to corner cases and the generation of previously unseen music attributes in a zero-shot setting.
To enhance the transparency of the conventional black-box generation process, we have devised a novel pipeline that furnishes procedural explainability and empowers researchers to conduct quality control at each step. This involves the emulation of the composition process followed by human experts and the decomposition of the generation task into four distinct stages.
(1) Conception Analysis: The Expert Module reconstructs the theme of the input text in terms of musical language. It identifies which elements of music composition are relevant to the text content and selects appropriate musical attributes.
(2) Draft Composition: Utilizing the chosen musical attributes as seeds, the Expert model employs various Composition Generator Modules to create a preliminary version of the piece.
(3) Self-Evaluation and Modification: The draft is subjected to the Voter Module where objective errors are identified based on music theory and subsequently corrected.
(4) Aesthetic Selection: Among all the error-free pieces, a subjective evaluation is conducted to select the composition deemed most aesthetically pleasing.

    Our contributions are manifold and intersect the domains of both music information retrieval and large language model applications:
\begin{itemize}
    \item  To our best knowledge, we are one of the first attempts to design an LLM agent as a melody composer. We develop a new agent architecture and melody composition pipeline including conceptual analysis, composition, self-evaluation, and aesthetic evaluation. 
    \item  Our system provides a "white-box" record of the music composition/modification process. Similar to a portfolio, it explains the AI's creative motivations and processes, thereby enriching the information space around symbolic compositions.
    \item  We leverage the intent understanding and dialog capabilities of LLMs to provide interactive functionalities. We maintain a "State Memory Tree" and "Historical Dialogue Records" to preserve long-term memories of both the creative and dialogue processes. This enables the LLM to serve as an interactive "assistant" for human composers.
\end{itemize}


\section{Related Work}

\textbf{Interactive Music Generation.} Current interactive music generation are mainly built on language interactive. Inspired by the diffusion-based text-to-image models, researchers have implemented text-to-audio music generation with diffusion models \cite{schneider2023mo, huang2023noise2music, zhu2023ernie} on in-house text-audio datasets. Mulan \cite{huang2022mulan} train a joint embedding space linking text and music audio, making it possible to eliminate the need for labeled text-audio pairs, as demonstrated in MusicLM \cite{agostinelli2023musiclm} and MeLoDy \cite{lam2023efficient}.  

Text-to-symbol generation models, on the other hand, offer greater interpretability and flexibility as the symbolic output is easier to be analyzed and edited. 
The majority of existing approaches involve introducing specific attributes in music as an intermediate bridge and using algorithmically extracted attributes for self-supervised attribute-to-music training \cite{lu2023musecoco, von2022figaro} or executing supervised training based on paired attribute-music data \cite{zhang2020butter, lu2023musecoco} to build attribute-to-music generation models. 
The advanced language model GPT-4 \cite{OpenAI2023GPT4TR} has shown to be capable of generating ABC notation music from natural language inputs, yet with limited success in terms of harmony and complexity \cite{bubeck2023sparks}.

\textbf{Augmenting LM by Agent Design.} Language Models can be augmented with well-designed reasoning procedures and external tools. For strategies to improve reasoning skills, Chain-of-Thoughts(CoT) \cite{CoT} first demonstrates that a simple sentence can elicit LLM's reasoning ability. Then \citet{zhou2023leasttomost} break down a complex problem into a series of simpler subproblems and then solve them in sequence. 
Besides forward reasoning, \cite{reflexion, self-refine, refiner} introduced the “self-reflection” mechanism, where LMs provide feedback to their generation candidates to improve LMs' performance on specific tasks. 
Tree-of-Thoughts (ToT) \cite{ToT} integrates searching and reflection, where LMs perform deliberate decision making by considering multiple different reasoning paths and self-evaluating choices to decide the next course of action. \citet{toolformer, AugmentedLM} proposed the `Tool-Augmented LM', where a tool is an external module that is typically called using a rule or a special token. In this paradigm, LMs can access knowledge that is not necessarily stored in its weights, such as a piece of factual knowledge for AI4Science \cite{ChemCrow}. Furthermore, \citet{general-purpose, HuggingGPT} demonstrate that LMs can also be used as a general-purpose interface with models pre-trained on different modalities.
Although these augmenting methods have achieved great success in many fields, how to build efficient LM agents for melody composition is still under-explored. ByteComposer integrates multistep-reasoning, self-reflection and multimodal-tool usage, enabling the LLM to compose music more accurately.


\section{ByteComposer: an LLM-powered melody composition agent}
Breaking away from the opaque nature of traditional text-to-music models, ByteComposer systematically dissects the music generation task into four deliberate stages, emulating human creative processes. The structured pipeline unfolds as follows:

\begin{enumerate}
    \item \textbf{Conception Analysis}: The input textual theme is meticulously deconstructed and examined in musical terminology to ascertain the creative musical elements relevant to the text, and subsequently, appropriate musical attributes are selected.
    
    \item \textbf{Draft Composition}: With the derived musical attributes of the input text serving as a seed, compositional techniques are deployed to draft an initial musical piece.
    
    \item \textbf{Self-Evaluation and Modification}: The preliminary draft is subjected to a thorough self-assessment based on musical theory to identify and rectify any objective inaccuracies.
    
    \item \textbf{Aesthetic Selection}: Among all error-free compositions, a subjective evaluation is undertaken to choose the composition that most resonate with individual aesthetic predilections.
\end{enumerate}

\begin{algorithm}[th]
    \caption{ByteComposer's Creative Workflow}
    \begin{algorithmic}[1]
        \Procedure{ByteComposer}{input\_text}
            \State \textbf{Conception Analysis:} \Call{AnalyzeTheme}{input\_text}
            \State \textbf{Compositional Drafting:} \Call{ComposeDraft}{MusicalAttributes}
            \State \textbf{Self-Evaluation:} \Call{EvaluateDraft}{Draft}
            \State \textbf{Aesthetic Selection:} \Call{SelectBest}{AllDrafts}
        \EndProcedure
    \end{algorithmic}
\end{algorithm}

\begin{figure*}
        \centering
        \vspace{2pt}
        \includegraphics[width=0.95\textwidth]{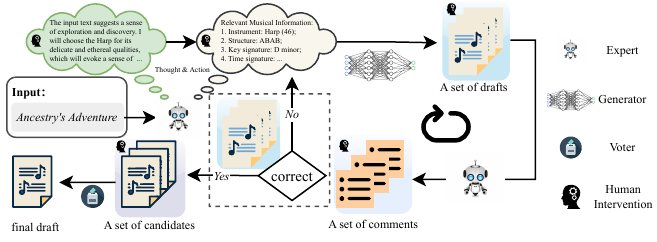}
        \vspace{-1em}
\caption{Overview of the ByteComposer System. From raw textual input, the system's Expert Module first analyzes emotional sentiment and thematic context, extracting primary features. This information is then passed to the Generator Module, which employs deep learning techniques to transform these textual features into initial musical motifs, leveraging knowledge from MIR. The Voter Module subsequently refines and evaluates the generated motifs, cross-referencing with historical compositions and utilizing real-time feedback loops. Conclusively, the Memory Module archives successful motifs, allowing the system to continuously learn and update its database, influencing future compositions. This seamless integration of modules ensures the creation of musically coherent and emotionally resonant pieces tailored to the input.}
        \label{fig:framework}
\end{figure*}

ByteComposer marshals three core modules---Expert, Generator, and Voter---to actualize its four-step composition process, as depicted in Figure 2. A cursory overview of these modules is proffered here.

To maintain a coherent continuum of the creative process and capture interaction data with users, ByteComposer is endowed with a Memory module. This module is architected to diligently record and store both the evolutionary trajectory of generated compositions and the discourse exchanges with users. The preservation of this data not only offers a historical narrative that can be referenced in future creative iterations but also cultivates a substantial repository for analyzing user interaction and feedback, thereby potentially honing the system's performance and user-centric adaptability over time. Core module is described as follows:

\begin{enumerate}
    \item \textbf{Expert:} Anchored by a Large Language Model (LLM), the Expert module navigates conceptual analysis and self-evaluation. Harnessing the LLM's general understanding, reasoning prowess, and music theory acumen, it translates users' queries into music descriptions for the Generator and provides insightful feedback for self-reflection.
    
    \item \textbf{Generator:} Entrusted with the composition task, the Generator module primarily concentrates on "attribute-to-music" generation and local problem rectification. Our empirical scrutiny juxtaposed LLM-based generation schemes against independent generation models, revealing a superior output quality in the latter.
    
    \item \textbf{Voter:} Recognizing the subjective essence of musical compositions, ByteComposer assimilates a Voter module to evaluate multiple candidate compositions, eventually selecting the one most congruent with human aesthetic discernment.
    
    \item \textbf{Memory:} Additionally, a Memory module is integrated to sift through and archive both intermediary texts and generated compositions throughout the entire creative journey.
\end{enumerate}


\subsection{Prompt Design}

Although LLMs have been demonstrated to be powerful few-shot learners across various tasks, their capabilities can only be fully unleashed through carefully designed prompts tailored to the desired behavior. The quality of the generated output is often sensitive to design choices in the prompt, sometimes down to the selection of punctuation marks. To this end, we identify core components of the prompts and evaluate different design choices through cross-validation. We ultimately classify prompts into two categories: process-related prompts and music theory-related prompts.

\begin{enumerate}
    \item \textbf{Process-Related Prompts:} To ensure the smooth operation of the entire workflow, the first step involves the design of appropriate process prompts. Clear and logical context cues guide the LLM to understand its current state and complete tasks within the current workflow accordingly.
    
    \item \textbf{Music Theory-Related Prompts:} To stimulate the LLM's music theory knowledge and enhance the quality of generation, we design prompts specifically related to music theory:
    \begin{enumerate}
        \item \textbf{Music Theory Explanation:} Explicit explanations of music theory attributes are included within the prompt.
        
        \item \textbf{Music Attribute Guidance:} By listing musical attributes, the LLM is guided to select those most congruent with the theme of the generation.
        
        \item \textbf{CoT and Few-Shot Learning:} Specific examples are provided for the LLM to reference.
    \end{enumerate}
\end{enumerate}

\subsection{Expert Module}


    
    
    


The Expert module is an integral part of the ByteComposer system, assuming a critical role in guiding the music composition process. This module is adept at analyzing user inputs and determining appropriate musical attributes for composition. Moreover, it seamlessly integrates with various music generation models to produce compositions that meet the user's specifications. A unique feature of this module is its ability to evaluate the quality and artistic merit of compositions, leveraging both objective criteria and subjective feedback. Additionally, it interacts closely with the Memory module, ensuring smooth transitions between system states and making decisions about progressing or revisiting steps based on the composition's current status.

The implementation of the Expert module is built upon Large Language Models (LLMs) and the strategic crafting of prompts to elicit desired musical outputs. Among the LLMs employed with Few-Shot Prompting, offering nuanced musical interpretations. We extensively evaluated versions underpinned by ChatGPT-3.5, ChatGPT-4, and LLama2. Recognizing the LLama2 model's potential for adaptability, we employed Supervised Fine-Tuning (SFT) to better calibrate it to musical tasks. This was complemented by the generation of around 90,000 prompts tailored for training, ensuring that our LLMs were well-equipped to handle complex musical assignments. The additional application of SFT bolstered the LLMs' ability to produce music that closely followed the set guidelines and user intentions.

\subsection{Generator Module}

The Generator Module is central to ByteComposer's music creation capabilities, offering a tailored approach to musical composition. One of its standout features is the attribute-controlled generation, which allows ByteComposer to craft pieces according to specific musical attributes, be it a certain mood or rhythm. Additionally, ByteComposer is acutely aware of the intricacies in music; hence, it incorporates a bar-level score generation mechanism. This functionality not only brings a fine-tuned sense of detail but also ensures that minor corrections or rewritings can be implemented, preserving the coherence and beauty of the piece.

At the technological core of the Generator Module lies the integration of state-of-the-art models and pioneering methodologies. By harnessing the ABC notation generation capabilities of GPT4, ByteComposer can access an extensive musical repertoire, translating into diverse and sophisticated compositions. Parallelly, the TunesFormer-Plus is another crucial asset. It extends the foundational dataset from TunesFormer~\cite{wu2023tunesformer} by integrating additional musical attributes and utilizes GPT-2 as its backbone. The introduction of a specialized bar-level score generation module within TunesFormer Plus further boosts ByteComposer's capacity for nuanced adjustments, ensuring each piece strikes the perfect melody.

\begin{figure}
    \centering
    \includegraphics[width=0.4\textwidth]{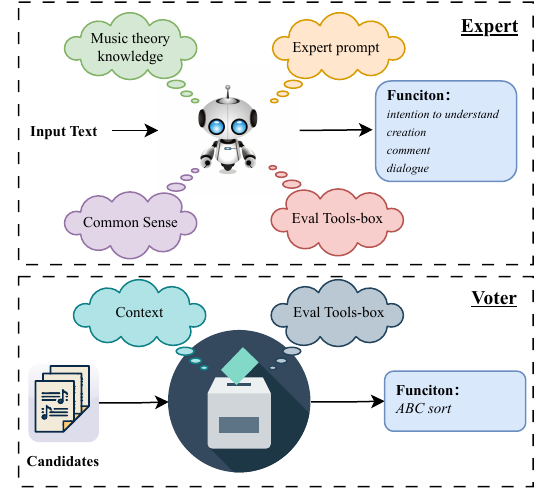}
    \caption{The interplay between the system's components. Input text is processed leveraging music theory knowledge, common sense, and context. This data then informs the Expert and Voter modules, facilitated by their respective evaluation toolboxes. The Expert module focuses on understanding and providing feedback on the creation, while the Voter module sorts the potential musical candidates.}
\end{figure}



\subsection{Voter And Memory Module}
\begin{figure}
    \centering
    \includegraphics[height=10cm]{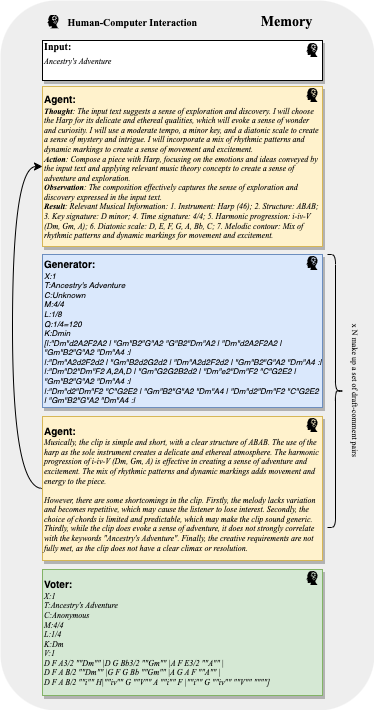}
    \caption{The Memory Module of ByteComposer.}
    \label{fig:mem}
\end{figure}

The Voter Module serves as a pivotal component in the ByteComposer pipeline, meticulously aligning generated compositions with human preferences. Its core objective is to sift through the compositions and select those that resonate most with human aesthetic sensibilities, especially when the generated outputs are devoid of any objective errors. The implementation strategy of the Voter Module employs the Llama2-QLoRA-SFT framework. In employing this framework, the Voter Module remains consistent with the RLHF training procedure, focusing its efforts on training solely the reward model and deliberately eschewing subsequent stages of PPO reinforcement learning\cite{christiano2017deep}. For training, a rich dataset was constituted, featuring 4,000 samples procured through GPT4, which were further enriched by an expert-annotated dataset of 1,000 target samples. This amalgamated dataset was then subjected to a combined training strategy, employing the strengths of Llama2 and QLoRA~\cite{dettmers2023qlora} to ensure the Voter Module's optimal performance.
To better model the composition reflection process, we learn from ToT\cite{ToT} and develop the State Memory Tree. A tree structure memory system to record the composition procedure and provide a reflective method to return to a previous node when facing difficulties in the creative process, as shown in Fig~\ref{fig:mem}.
Particularly, each node in this tree, structured as a tuple, captures the generation stage, context, and current text data. 
Besides reflection, the memory system can also work as clues for future similar compositions. ByteComposer retrieves from this tree using either breadth-first or depth-first search by state data. Alongside the tree, a message queue named "Historical Dialog Records" logs all user interactions.

\section{Datasets and Settings}
\subsection{Dataset}
\textbf{Music Generation Training Data.}
 The training dataset for our self-developed generator is identical to that used by TunesFormer~\cite{wu2023tunesformer}, namely, the Irish Massive ABC Notation dataset, which consists of 216,284 Irish tunes in ABC notation. 214,122 tunes for training, and 2,162 tunes for validation.
  

Custom attribute labels such as note density per measure, and average pitch fluctuation per measure were extracted using self-developed tools.\\

\textbf{Large Language Model Supervised Fine-Tuning Data.}
We assembled a dataset encompassing 2,128 professionally annotated musical queries. Leveraging expert-annotated seed queries, we adopted a self-instruction methodology to yield an extensive music domain dataset, comprising 91,341 entries, generated using GPT-4. This dataset spans a diverse range of facets including music theory elucidations, conceptual deliberations, score appraisals, and interpretations of musical intent. This augmented dataset was subsequently employed to fine-tune open-source large language models.
We curated a dataset containing 1,000 subjective evaluation entries through meticulous expert annotation. Each entry encapsulates a ranking of four ABC notation results. Through permutation and shuffling of these rankings, we engendered an expanded set of pairwise data, significantly bolstering the training and validation processes of our voter module.

\begin{table*}[!ht]\scriptsize
    \centering
    \caption{Large Language Model Supervised Fine-Tuning Data}
    \renewcommand\arraystretch{1.2}
    \begin{tabular}{c|c|c|c}
    \hline
        Type of Data & Application & Source & Number \\ \hline
        Basic music theory QA questions & Strengthen LLM music-related Q\&A skills & Expert collection & 2,128 \\ 
        Music theory conception & Comprehension and conception of input text & GPT4 generation & 30,000 \\ 
        Control code generation & Learn to use custom generative models & GPT4 generation & 15,000 \\ 
        Score evaluation opinions & Learn to check and evaluate music scores using tool tools & GPT4 generation & 30,000 \\ 
        Designated plan & Analyze current status and plan next steps & GPT4 generation & 6,131 \\ \hline
    \end{tabular}
    \label{table1}
\end{table*}

\subsection{Implementation Details}
For the experiments, we employed the OpenAI GPT4 (\texttt{gpt-4-32k}) API and GPT3.5 (\texttt{gpt-3.5-turbo-16k}) API models, utilizing Python 3.9 to execute the generated programs. Additionally, we explored the llama2-70B/13B/7B as potential base models for our open-source model, eventually settling on llama2-70B.

We configured the model with LoRA rank parameter \texttt{lora\_r = 64}, scaling parameter \texttt{lora\_alpha = 16}, and dropout probability \texttt{lora\_dropout = 0.1}. 4-bit precision base model loading was activated (\texttt{use\_4bit = True}) with compute data type \texttt{bnb\_4bit\_compute\_dtype = "float16"} and quantization type \texttt{bnb\_4bit\_quant\_type = "nf4"}. Nested quantization was not activated (\texttt{use\_nested\_quant = False}).

     The model was trained on NVIDIA A100-SXM4-80GB x 8 for 4 epochs, completing within an approximate timeframe of 3 days.

Our generation model drew inspiration from Tunesformer, adopting the GPT-2 architecture to align experimental standards with Tunesformer parameters. We extended the model by introducing two additional attributes -- velocity, note density, and curvature -- to control rhythm and melody fluctuations. The enhanced model, termed Tunesformer-Plus, partitioned the training set into 2-section units, generating analogous 2-measure segments to facilitate fragment-level modifications.

\begin{figure}
    \centering
    \includegraphics{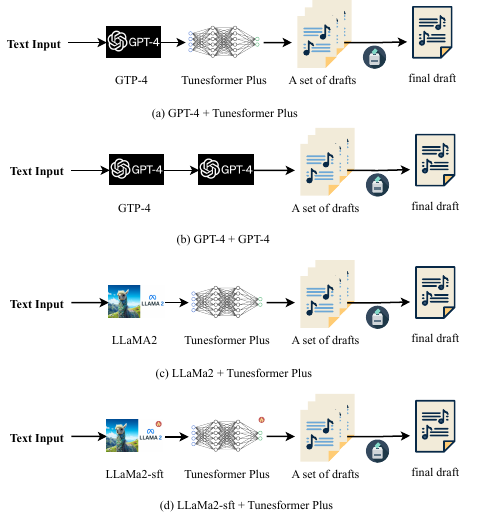}
    \caption{Diverse agent configurations are enabled through the combination of different functional modules, further augmented by the support for custom model module components. This modularity not only facilitates a tailored approach to specific tasks but also enhances the system's adaptability to evolving requirements and novel applications.}
\end{figure}

\subsection{Agent System Configuration}
We employed various combinations to implement the ByteComposer architecture, the details of which can be seen in Figure 4. Through ablation experiments, the optimal setup for ByteComposer was found, with the open-source LLM llama2 size set at 70B.
The ABC generation quality of ChatGPT lagged significantly behind GPT4, while llama2-based models struggled with ABC notation generation. So GPT4 core-based ByteComposer schemes, two configurations were tested: one with Tunesformer-plus and another with GPT4. In open-source configurations, the Llama2 core with GPT4 abc generator combination was omitted to maintain opensource consistency.
Notably, both the LLM and generator in our ByteComposer architecture are interchangeable. Other symbolic generation models could replace the generator, even those based on MIDI or audio. For non-ABC music mediums, corresponding perceptual components need to be developed for metric quantification and feedback to the LLM for further processing.

\section{Experiments and Evaluation}

\subsection{Metrics}
        \textbf{Time Signature Error Rate(TSER).} This metric evaluates whether the time signature in the ABC notation header corresponds to the beat count in each measure. An error is flagged if any measure has an incorrect beat count.\\
        \textbf{Instrument Range Error Rate(IRER).} Every instrument has a designated playable range. Using an expert-defined instrument range table, we check if the generated notes exceed this range, marking any note outside the range as an error. \\
        \textbf{Score Information Completeness Rate(SICR).} A complete ABC score should have title, tempo, etc. we have developed code to quantify the completeness of the ABC notation. A score is deemed incomplete if any of these attributes are missing.\\
        \textbf{Average Attribute Accuracy(AAA).} This metric evaluates if the attributes in the generated ABC notation align with those conceived during the designing phase. We calculate the accuracy for each musical attribute and then compute their average.
        \textbf{Voting Accuracy(VA).} This metric evaluates the ability of the voting model to correctly choose the better ABC notation candidate, as labeled by humans, from two given options.

    \subsection{Objective Experiments}
    The objective experiment primarily examines agent-generated music scores, with a focus on objective metrics like score completeness and attribute control accuracy. To assess and quantify these metrics, we utilize a suite of tools we term as "Eval Tools-box", comprising the open-source music21 library and our custom ABC parsing tool, to extract pertinent labels from ABC notation instances.\\
    We draw comparisons with two of the most powerful chatbots currently available, and also benchmark against a state-of-the-art ABC notation music composition models in terms of score generation metrics. From Table 2, we can find that the symbol music generation capabilities of GPT4 and Chatgpt are relatively poor. The simple generation model TunesFormer has a better generation effect, but the completeness of the chart information is relatively poor. At the same time, because we expanded the attribute part, we did not calculate the attribute control indicators of Tunesformer. Finally, we utilized the ByteComposer framework to enhance the comprehensive capabilities of both models significantly.

\begin{table}[!ht]\scriptsize
    \centering
    \caption{Objective Experiments Scores}
    \renewcommand\arraystretch{1.2}
    \begin{tabular}{c|*{5}{p{0.45cm}}}
        \hline
        Model & TSER$\downarrow$ & IRER$\downarrow$ & SICR$\uparrow$ & AAA$\uparrow$ & VA$\uparrow$ \\ \hline
        ChatGPT & 89.8\% & 67.7\% & 40.3\% & 37.5\% & 52.8\% \\ 
        GPT4 & 68.8\% & 48.8\% & 56.3\% & 56.3\% & 60.3\% \\ 
        TunesFormer & 11.3\% & - & - & - & - \\ 
        ByteComposer-GPT4 & \textbf{1.8}\% & \textbf{19.8}\% & \textbf{83.4}\% & \textbf{81.3}\% & 64.3\% \\ 
        ByteComposer-Llama2-SFT & 2.6\% & 21.6\% & 79.3\% & 80.4\% & \textbf{77.1}\% \\ \hline
    \end{tabular}
    \label{table2}
\end{table}

\begin{table*}[!ht]\scriptsize
    \centering
    \caption{Evaluations of the Competency of LLMs in Music Theory Interaction.  }
    \renewcommand\arraystretch{1.2}
    \resizebox{1.5\columnwidth}{!}{
    \begin{tabular}{c|cccc}
    \hline
        Model & {Accuracy $\uparrow$} & {Text Fluency Score $\uparrow$} & {Comprehension Score $\uparrow$} & {Problem Resolution Rate $\uparrow$} \\ \hline
        {Llama2-70B-SFT(Our proposed)} & \textbf{88.06\%} & \textbf{96.94\%} & \textbf{96.73\%} & 51.50\% \\ 
        {Llama2-13B-chat} & 85.82\% & 94.8\% & 89.8\% & 36.00\% \\ 
        {Falcon-40B} & 86.73\% & 93.78\% & 89.8\% & -\\ 
        {Llama2-70B-chat} & 87.96\% & 95.71\% & 90.82\% & 37.00\% \\ 
        {WizardLM-70B-V1.0} & 80.61\% & 96.12\% & 85\% & 35.50\%\\ 
        {Platypus2-70B-instruct} & 77.65\% & 82.55\% & 83.06\% & -\\ \hline
        {GPT-4} & \textbf{93.16\%} & \textbf{99.31\%} & \textbf{99.83\%} & \textbf{63.50\%} \\ \hline
    \end{tabular}}
    \label{table3}
\end{table*}

\begin{table}[!ht]\tiny
    \centering
    \caption{Generation Quality Evaluation}
    \renewcommand\arraystretch{1.4}
    \resizebox{1\columnwidth}{!}{
    \begin{tabular}{c|cc}
    \hline
        Model & Music Generation Quality$\uparrow$ & Music Conception Ability$\uparrow$ \\ \hline
        ChatGPT & 1.21 & 3.42 \\ \hline
        GPT4 & 2.32 & \textbf{4.06} \\ \hline
        MuseCoco & 3.53  & - \\ \hline
        ByteComposer-GPT4 & \textbf{4.42} & \textbf{4.08} \\ \hline
        ByteComposer-Llama2-SFT & \textbf{4.31} & 3.82 \\ \hline
    \end{tabular}
    }
    \label{table4}
     \vspace{-1em}
\end{table}

\begin{table}[!ht]\scriptsize
    \centering
    \caption{Ablation Experiments}
    \renewcommand\arraystretch{1.2}
    \resizebox{1\columnwidth}{!}{
    \begin{tabular}{c|cc}
    \hline
        Model & Music Generation Quality$\uparrow$ & Music Conception Ability$\uparrow$ \\ \hline
        \makecell[tl]{GPT4 }& 2.32 & 4.06 \\   \hline
        \makecell[tl]{GPT4 + CoT prompt} & 2.85 & 4.12  \\   \hline
        \makecell[tl]{GPT4 + CoT prompt \\ + Tuensformer-Plus} & \makecell{\vspace{-1.1em}3.87} & \makecell{\vspace{-1.1em}4.12}  \\  \hline
        \makecell[tl]{GPT4 + CoT prompt + \\ Tuensformer-Plus + Self-Eval} & \makecell{\vspace{-1.1em}4.21} & \makecell{\vspace{-1.1em}4.08} \\  \hline
        \makecell[tl]{GPT4 + CoT prompt + \\ Tuensformer-Plus + Self-Eval + \\ Voter (ByteComposer-GPT4)} &   \makecell{\vspace{-2em}\textbf{4.42}}  & \makecell{\vspace{-2em}\textbf{4.08}} \\ \hline
    \end{tabular}
    }
    \label{table5}
    \vspace{-1em}
\end{table}

\subsection{Expert Subjective Experiments}
    \subsubsection{Participant Information}
    To rigorously evaluate the musical capabilities of ByteComposer from a music expert's perspective, we detailed the participant information and evaluation process as follows: 
Invited 15 volunteers, all interested in AI-based music production, and working in the field of music and audio technology or production, being professional musicians. Professionals evaluated the outputs under the same playback environment and model inference parameters, in a blind-review setup.\\
\textbf{Experience in Music Production.} 40\% had between 2 to 5 years of experience, while 60\% possessed more than 5 years of experience.\\
\textbf{Experience in Music Performance.} All had more than 10 years of experience.\\

\subsubsection{Competency of LLMs in Music Theory Interaction} 
In this chapter, our objective is to thoroughly test the LLM's musical and interactive capabilities, to ensure the assembled agent possesses the ability for music theory inquiries and normal interactions. \\
\textbf{Music Theory Comprehension.}
In the first round, we evaluated several open-source LLMs, as well as GPT4 for music theory comprehension evaluation help us judge the capability of the open source models we trained. Human Experts created an 500 professional questions from various music professional latitudes, and scored each model result from three dimensions (Accuracy, Text Fluency, Comprehension of Questions).\\
\textbf{Music Composition Facilitation.}
Through the first round of evaluation, we filtered out some models with lower scores, and the remaining ones were evaluated in the second round. Users engage in conversations related to music composition with LLM, and if the responses satisfied the users, it indicated problem resolution. We set up 500 user queries and scored the results of each model. \\
\textbf{Experiments Conclusion.}
As can be seen from Table 3, The model we trained through fine-tuning, namely Llama2-70B-SFT, is stronger than other open source models in terms of professional capabilities. The main reason should be the results of fine-tuning on music field data and professional question and answer data. However, the comprehensive indicators are still behind GPT4.
\vspace{-0.8em}           
\subsubsection{Generation Quality Evaluation}
Evaluation Metrics include Music Generation Quality examining Theme Fidelity, Music Quality, and Score Reasonability, along with Music Conception Ability examining Reasonability of Conception and Music Analysis Evaluation Ability. Comparative Schemes involve evaluating generation quality across different methods like ChatGPT, GPT4, ByteComposer variants, and MuseCoco\cite{lu2023musecoco} against 200 keywords, with experts scoring the results on a 0-5 scale. As shown in Table 4, the conclusion reveals that the generation quality of MuseCoco is pleasing, but the music symbol generation model has no ability to explain the conception of musical scores, whereas ChatGPT and GPT4 excel in process description but fall short in  generation quality. ByteComposer-based solutions exhibit competitive scores in both generation quality and process description. \\
\subsection{Professional Analysis and Summary}
From a musical creation conceptualization perspective, the agent demonstrates a relatively complete music design ability (Thought-Action-Observation-Result). Initially, it can capture the underlying content and emotions from keywords, abstracting corresponding musical features like instrument and mode selection reflecting respective musical characters, and rhythm and tempo choices depicting different musical emotions. Subsequently, these refined musical features are combined to form a relatively complete musical concept. While the information extracted from text is quite accurate, the creative methods employed are still limited and simplistic. 
On the other hand, the model possesses commendable music analysis ability, capable of describing the song structure, mode, rhythm, harmony, and melodic patterns. However, the depth and detail of analysis are yet to match top professional music analysis standards, and there is room for improvement. Noteworthy is the model's capability to provide modification suggestions while evaluating music, with accurate directions, although lacking in diversity. Overall, the current AI composer has reached the level of a beginner composer.



\subsection{Ablation Experiments}

During the Generation Quality Evaluation experiments, we compared module ablation results longitudinally. The overall results are shown in Table 5. Baseline for these experiments is GPT4, which directly generates ABC notation music from user prompts. We introduced expert-designed CoT prompts, which improved generation quality and achieved the highest Music Conception Ability scores. However, experts clearly favor music generated by combining GPT4 with the music generation model Tunesformer-Plus. Additionally, we found that using the "self-eval" module, representing the self-modification phase (the third stage), enhances Music Generation Quality but results in a slight decrease in Music Conception Ability. Our observations suggest this decrease is because experts tend to assign higher scores when the model accurately evaluates low-quality music. Finally, the "voter" module enhances music quality by selecting more preferred music pieces. The system hyperparameters and Llama2-SFT hyperparameter experiments are provided in the appendix.

\section{Conclusion}
In conclusion, we have demonstrated that music composition agents based on Large Language Models (LLMs) can serve as professional human composers by your side, capable of interactive engagement, with the generated symbolic works being more comprehensible and meaningful to individuals. Concurrently, we have implemented ByteComposer-Llama2-SFT using open-source solutions, proving its commendable effectiveness. Through the creative architecture of ByteComposer, the potential of large language models is effectively harnessed, amplifying the value of existing generation models. 

\nocite{*}
\newpage
\section{Bibliographical References}\label{sec:reference}

\bibliographystyle{lrec-coling2024-natbib}
\bibliography{lrec-coling2024-example}


\end{document}